# Development and Mass Production of a Mixture of LAB- and DIN-based Gadolinium-loaded Liquid Scintillator for the NEOS Short-baseline Neutrino Experiment

Ba Ro Kim[2], Boyoung Han[5], Eun-ju Jeon[4], Kyung Kwang Joo[2]*, H. J. Kim[6], Hyunsoo Kim[7], Jinyu Kim[7], Yeongduk Kim[4], Youngju Ko[3], Jaison Lee[4], Jooyoung Lee[6], Moohyun Lee[4], Kyungju Ma[7], Yoomin Oh[4], Hyangkyu Park[4], Kang-soon Park[4], Kyungmin Seo[1], Gwang-Min Seon[5], Kim Siyeon[3]

[1] *Department of Physics, Chonbuk National University, Jeonju, 54896, Korea*

[2] *Institute for Universe & Elementary Particels, Department of Physics, Chonnam National University, Gwangju, 61186, Korea*

[3] *Department of Physics, Chung Ang University, Seoul, 06911, Korea*

[4] *Center for Underground Physics, Institute of Basic Science, Daejeon, 34057, Korea*

[5] *Neutron Science Division, Korea Atomic Energy Research Institute, Daejeon, 34057, Korea*

[6] *Department of Physics, Kyungpook National University, Daegu, 41566, Korea*

[7] *Department of Physics & Astronomy, Sejong University, Seoul, 05006, Korea*

Correspondence should be addressed to Kyung Kwang Joo; kkjoo@chonnam.ac.kr

**Abstract**

A new experiment, which is called as NEOS (NEutrino Oscillation at Short baseline), is proposed on the site of Hanbit reactors at Yonggwang, South Korea, to investigate a reactor antineutrino anomaly. A homogeneous NEOS detector having a 1000-ℓ target volume has been constructed and deployed at the tendon gallery ~25 m away from the reactor core. A linear alkylbenzene (LAB) is used as a main base solvent of the NEOS detector. Furthermore, a di-isopropylnaphthalene (DIN) is added to improve the light output and pulse shape discrimination (PSD) ability. The ratio of LAB to DIN is 90:10. PPO (3 g/ℓ) and bis-MSB (30 mg/ℓ) are dissolved to formulate the mixture of LAB- and DIN-based liquid scintillator (LS). Then, ~0.5% gadolinium (Gd) is loaded into the LS by using the solvent-solvent extraction technique. In this paper, we report the characteristics of Gd-loaded LS (GdLS) for the NEOS detector and the handling during mass production.

## 1. Introduction

The NEOS (NEutrino Oscillation at Short baseline) experiment is a reactor-based short-baseline neutrino oscillation experiment to investigate a reactor antineutrino anomaly by using electron antineutrinos emitted from the Hanbit nuclear power plant in South Korea. A NEOS detector is installed on the concrete foundation of the tendon gallery of the Hanbit reactor number 4 building, in a dedicated space beneath the reactor core, as shown in Figure 1. The NEOS detector is a 1000-ℓ homogeneous detector using a well-known liquid scintillator (LS) technology. A linear alkylbenzene (LAB, $C_nH_{2n+1}$-$C_6H_5$, where n = 10~13) is used as the main base organic solvent of the NEOS detector. Small portion of di-isopropylnaphthalene (DIN, $C_{16}H_{20}$) is added to improve the light output capability and pulse shape discrimination (PSD).

Reactor antineutrinos are detected through the inverse beta decay (IBD, $\bar{\nu}_e + p \rightarrow e^+ + n$) reaction, followed by neutron capture [1, 2, 3]. The prompt positron annihilates and yields 1~8 MeV of visible energy. If the neutron is captured by a hydrogen, it gives off a gamma with an energy of 2.2 MeV within ~200 μs. However, if the LS is loaded with a small amount of gadolinium (~0.1% Gd), having a large neutron capture cross section, the delayed neutron capture signal can produce several gammas whose total energy amounts to ~8 MeV within ~30 μs [4, 5]. This energy is well above the low energy background.

During the 80's and 90's, several short-baseline experiments were carried out with a detector located a few tens of meters away from reactor cores [6, 7, 8]. In these experiments, $R = N_o/N_p$ was calculated, where $N_o$ represents the number of observed events, and $N_p$ represents the number of predicted events [9, 10, 11]. Recently a more refined evaluation of the predicted neutrino event rate has been developed by several people [12, 13, 14]. They claimed that the value of $R$ shifts to below unity, which is a condition that has been termed "a reactor antineutrino anomaly". Currently, several short-baseline reactor experiments are proposed to investigate the anomaly [15, 16].

We develop and synthesize a mixture of LAB- and DIN-based Gd-loaded LS for the NEOS experiment. Several optical and physical properties, such as transmittance, absorption, light yield (LY), pulse shape discrimination (PSD) separation capability, Gd concentration, and water concentration, are measured for quality control. In this paper, we briefly report the characteristics and the handling of the mass production of the GdLS for the NEOS detector [17].

**2. Development and Mass Production of the Mixture of LAB- and DIN-based Gd-loaded LS**

*2.1 Mixture of LAB- and DIN-based Gd-loaded liquid scintillator*

A liquid scintillator (LS) detector is widely used to detect neutrinos emitted from a reactor [1, 2, 3]. A liquid scintillator consists of a base solvent and a fluor. A secondary wavelength shifter (WLS) can be used to match the light emission to the spectral response of the PMT. Furthermore, in order to enhance the efficiency of the neutron capture signal, metal, such as gadolinium (Gd), can be loaded into the LS.

For an organic base solvent, LAB is used for the NEOS detector. LAB is known to have high transmittance and a long attenuation length [18, 19]. Furthermore, LAB has a high flashpoint (140ºC) and poses no risk of harming the human body or the environment. These factors alleviate safety concerns for LAB use at strictly regulated reactor power plant sites. The purchased LAB is passed through a 0.1-μm pore size Teflon membrane filter to remove small particulates. For the fluor, 3 g/ℓ of PPO (2,5 diphenyloxazole) is used to get light emission. As a WLS, 30 mg/ℓ of bis-MSB (1,4-bis(2-methylstyryl)benzene) is dissolved to shift wavelengths to the sensitive region of the Hamamatsu R5912 PMT. These quantities of mixture are evaluated as having the best optical and chemical properties for the experiment.

In the case of DIN, a commercially available DIN-based LS, such as the Ultima Gold-F (UG-F), was purchased from Perkin Elmer. The attenuation length of the DIN-based LS is shorter than that of LAB-based LS [20]. However, in terms of pulse shape discrimination (PSD) ability, DIN-based LS is generally much better than LAB-based LS. PSD can be affected by selecting different organic base solvents. Therefore we decide to mix UG-F into a LAB-based GdLS.

All of the equipment installed for the mass production are wiped down in the clean room, and then bring into the main production line. During all synthesis processes in the mass production of loading Gd into LS, much water is needed. Therefore, a water plant with a purification and circulation system is installed to produce ~20 liters of 16-MΩ high-purity water per hour. All materials in contact with the liquids during the mass production of GdLS are glass, acrylic, or Teflon-coated.

In order to load the Gd into the liquid scintillator, the Gd is dissolved in the organic liquid in the form of a Gd-carboxylate complex. For synthesizing LAB-based GdLS, we use a well-established solvent-solvent extraction technique involving the following two chemical reactions [21]:

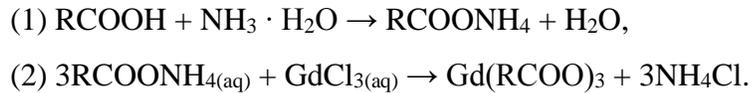

(1) $RCOOH + NH_3 \cdot H_2O \rightarrow RCOONH_4 + H_2O$,
(2) $3RCOONH_{4(aq)} + GdCl_{3(aq)} \rightarrow Gd(RCOO)_3 + 3NH_4Cl$.

The first step is to neutralize the reaction process of the carboxylic acid (TMHA, 3,5,5-trimethylhexanoic acid) with ammonium hydroxide. The second is to produce a mixture of two aqueous solutions, which leads to the formation of a Gd-carboxylate complex. All reactions are carried out in an organic solvent, LAB. The reactions are very sensitive to pH. The solubility of the organic metal is higher in the organic solvent than in water. After the reactions, the organic solvent and water are separable due to the difference of density between oil and water. However, much skill and effort are needed to remove the water for the final gadolinium-loaded liquid scintillator.

*2.2 Mass production of mixture of Gd-loaded LAB-based LS and DIN-based LS*

The mass production system should handle approximately a total of ~1300 ℓ of GdLS. Even though the NEOS detector needs 1000 ℓ of GdLS, ~300 ℓ more is produced for quality control, long-term stability checks, etc. The mass production system consists of high-density polyethylene (HDPE) storage tanks, acrylic mixing vessels including membrane filters, a mechanical pump, pneumatic values, Teflon hose pipe line, a nitrogen ($N_2$) purging device, a swage system, and the aforementioned water production system. Several specialized acrylic tanks are designed to carry out all of the fluid operations for the mass production. The mass production and mixing scheme is briefly shown in Figure 2. Three stages of synthesis are performed, based on the acrylic tank sizes of mass production systems in each stage.

In stage A, the first, the master Gd-LAB solution with ~0.8% Gd high concentration is made in 10 batches. With each batch, we synthesized ~90 ℓ of the master Gd-LAB solution and then put it into the ~1500-ℓ HDPE storage tank, until ten batches are accumulated in the tank. A total of 900 ℓ of the master Gd-LAB solution are produced in stage A. In stage B, 270 ℓ of 0.1-micron filtered purified LAB is added into a mixing storage tank and stirred with an adequate impeller speed. During stage C, PPO and bis-MSB and 150 ℓ of UG-F are added into the storage tank and mixed together. The ratio of the LAB-based Gd LS and UG-F for the NEOS liquid scintillator is 90:10. Finally, the Gd concentration is diluted to the ~0.5% level, which is our original goal.

If oxygen is present in the LS, the light output and PSD are strongly affected [22]. Therefore, it is essential to remove the oxygen. To this end, a gas operating system is implemented to handle the use of nitrogen gas. Nitrogen purging is done thoroughly at the end of each stage.

## 3. Measurements of Optical Properties of Gd-loaded LS

*3.1 Light yield (LY)*

Sufficient and consistent light output of each synthesized batch is essential for the mass production. After synthesizing each ~90-ℓ batch of Gd-LAB, a ~300-mℓ Gd-LAB sample is extracted for quality control. An LS master solution, which PPO and bis-MSB are dissolved at high concentration, is added into a small portion of the extracted Gd-LAB sample to check LY quickly before moving on to the next steps. The light output of this small GdLS sample is checked in two ways. First, visual inspection is performed. Each batch is illuminated with an ultraviolet (~250 nm) lamp. Strong scintillating light emitted from each batch is clearly seen. To perform a more detailed inspection, the energy spectrum is determined using a $^{137}$Cs gamma-ray source. The energy spectra of each batch are analyzed and fitted by using the proper function for determining the position of the peak from the Compton scattering edge. The full width at half maximum (FWHM) is used to find the peak of the distribution. The results are shown in Figure 3 and compared with some other reference samples, such as LAB-based LS with 3 g/ℓ of PPO and 30 mg/ℓ bis-MSB dissolved within, and the UG-F sample itself. As expected, the LY of each GdLS sample is the same, within the uncertainty and less than that of LS. The UG-F sample have the highest LY value and the light output was ~50% greater than that of the LAB-based Gd-loaded LS at the end of stage A. Because UG-F is added in stage C, the LY of the final mixture of LAB- and UG-F-based GdLS after stage C is found to be higher than that of GdLS sample.

*3.2 Gd concentration*

Because the Gd concentration is directly related to the capture time of the neutron, it is important to know the exact concentration of Gd. In order to extract Gd from the gadolinium-loaded liquid scintillator, an complexometric EDTA (Ethylene Diamine Tetra Acetic acid) titration method with xylenol orange as indicator is applied [23, 24]. For the calibration of the Gd concentration, certified 0.1%, 0.3%, 0.5%, 0.7%, and 1% standard solutions are used. After careful calibrations, the values of Gd concentration in the A and B stages are 0.75 and 0.62%, respectively. Then finally it is diluted to 0.54% by adding UG-F in stage C, and the NEOS detector is filled with it. The uncertainty level of EDTA method is ±0.03%.

*3.3 Densities*

A reasonably good density measurement is necessary to determine the number of free protons in the target of the NEOS detector. The density at each stage is measured using a portable density meter (DA-130N, KEM) with a resolution of 0.001 g/cm$^3$ at 0~40°C [25], which is sufficient for our purposes. The measured densities are 0.862, 0.859, and 0.870 g/cm$^3$, respectively. For reference, the density of the pure LAB at 28°C is measured with values of 0.852 ± 0.001 g/cm$^3$.

*3.4 Pulse shape discrimination (PSD)*

In the aboveground environment, which has much external background radiation, having the PSD ability of GdLS is important for particle identification. It is possible to distinguish between a fast neutron event and a γ–ray event by obtaining the ratio between the total charge ($Q_{\text{total}}$) and the tail charge ($Q_{\text{tail}}$) of the pulse, $\frac{Q_{\text{tail}}}{Q_{\text{total}}}$ [26, 27, 28]. The $Q_{\text{tail}}$ value is defined as the tail charge part over some duration after the maximum pulse height is reached. The $Q_{\text{tail}}$ and $Q_{\text{total}}$ are measured by using a $^{252}$Cf source that emits both γ–rays and fast neutrons with a mean energy of 2.14 MeV. In addition, the quality of PSD can be expressed in terms of the figure of merit (FoM = $\Delta S/\sqrt{\sigma_1^2 + \sigma_2^2}$), where $\Delta S$ represents the distance between the peaks of the two distributions, and $\sigma_1$ and $\sigma_2$ are their respective standard deviations [26]. We take the maximum FoM value by applying different high voltages (HV). It should be noted that by adding a small portion of the UG-F into the LAB-based GdLS sample, the light yield and the FoM is significantly enhanced. According to our previous study, the light yield (LY) is enhanced by about 30%, and the FoM is drastically improved by more than ~80% [17]. As seen in Figure 4, a relative discrimination of neutron (n) and γ–ray signals can be obtained.

*3.5 Water content*

If water exists in the GdLS, the stability of a GdLS will be greatly affected [29, 30, 31]. Due to the nature of the synthesizing process currently we used, water creation cannot be avoided. Even though we attempt to remove the water from each batch as much as possible, a small portion of water might still exist in the GdLS. At each stage, the water content is measured by using Karl-Fischer titration, which is a classic titration method of analytic chemistry that provides a high level of accuracy. A Mettler Toledo C20X Karl-Fischer coulometer with a resolution of 0.1 μg is used. The values of the water content at stage A and B are found to be 650 and 500 ± 5.0 ppm, respectively.

The water content in the final GdLS as a function of the quantity of $N_2$ flowed are shown in Figure 5. The amount of gas needed to flush the 1000 ℓ of GdLS can be estimated using the simple expression $Q = Q_i e^{-\lambda N}$, where $\lambda$ is constant related to a relaxation time, $Q_i$ is the initial concentration of water in ppm and N is the nitrogen gas ($N_2$) quantity injected into the GdLS. To achieve ~250 ppm level of water concentration from ~500 ppm, ~400,000 liters of nitrogen gas at 1 atm is needed. The water content decreases and then become stable at a value of ~200. 0 ± 5.0 ppm within our specification.

*3.6 Transmittance*

According to the past reactor neutrino experiments, for a successful experiment the GdLS should be optically transparent for several years [30, 32]. Absorption and transmittance spectra of NEOS GdLS are measured. The light transmittance value T is defined as T= $I/I_0$, where $I_0$ and I are the intensities of the incident and the emitted light, respectively [32, 33, 34]. The measurement is carried out in a 10-cm-path-length cylindrical quartz cuvette cell. The transmittance value is measured by using a Shimadzu UV/Vis 1800 spectrophotometer and is scanned over the wavelength range of 300-500 nm. The absorption value *A* is related to T as *A*=−log(T). The spectrum is related to the energy transfer mechanism between absorption and emission. As shown in Figure 6, the transmittance value of the NEOS liquid scintillator is over ~98% in the wavelength region from 420 nm to 500 nm. This is the wavelength region where the main scintillation occurs, and our PMT is the most sensitive. UG-F and LAB-based Gd-unloaded LS are used for comparison. As we pointed out, the transmittance of UG-F itself is lower than that of LAB-based LS. However, adding 90% of LAB into 10% of UG-F enhances the transmittance value up to ~98%.

## 4. Summary

The NEOS detector is specifically built to yield a reliable measurement of the reactor antineutrino anomaly in the vicinity of operating Hanbit nuclear reactor cores that have a thermal power output of 16.4 GW$_{th}$. The NEOS detector uses ~0.5% Gd-loaded LS, based on the mixed solvents of LAB and UG-F. Adding 10% of the UG-F into the LAB-based 0.5% GdLS improves the light yield, the FoM of PSD, and the transmittance. This mixture shows promising results for use in the short-baseline NEOS reactor neutrino experiment. The 1000 ℓ of GdLS is synthesized using a solvent-solvent extraction method, which is robust and reproducible during the entire production period. Quality controls were applied at each stage of the mixing procedure, such as checks on the light yield, Gd concentration, transmittance, density and water contents. All physical and optical properties are satisfactory. The 1000 ℓ of GdLS is transported to the location, and the homogeneous NEOS detector is filled with it and is in use for reactor antineutrino detection. The NEOS experiment started to take data, as of August 2015.

**Conflict of Interests**

The authors declare that there is no conflict of interests regarding the publication of this paper.

**Acknowledgements**

This work was supported by the Korea Neutrino Research Center, which was established with a grant from the National Research Foundation of Korea (NRF) funded by the Korean government (MSIP) (No. 2012-0001176), the Basic Science Research program through the National Research Foundation of Korea (NRF) funded by the Ministry of Education (2012M2B2A6030210), Samsung Science & Technology Foundation (SSTF-BA1402-06), Chonnam National University (2014), and Grant No. IBS-R016-D1.

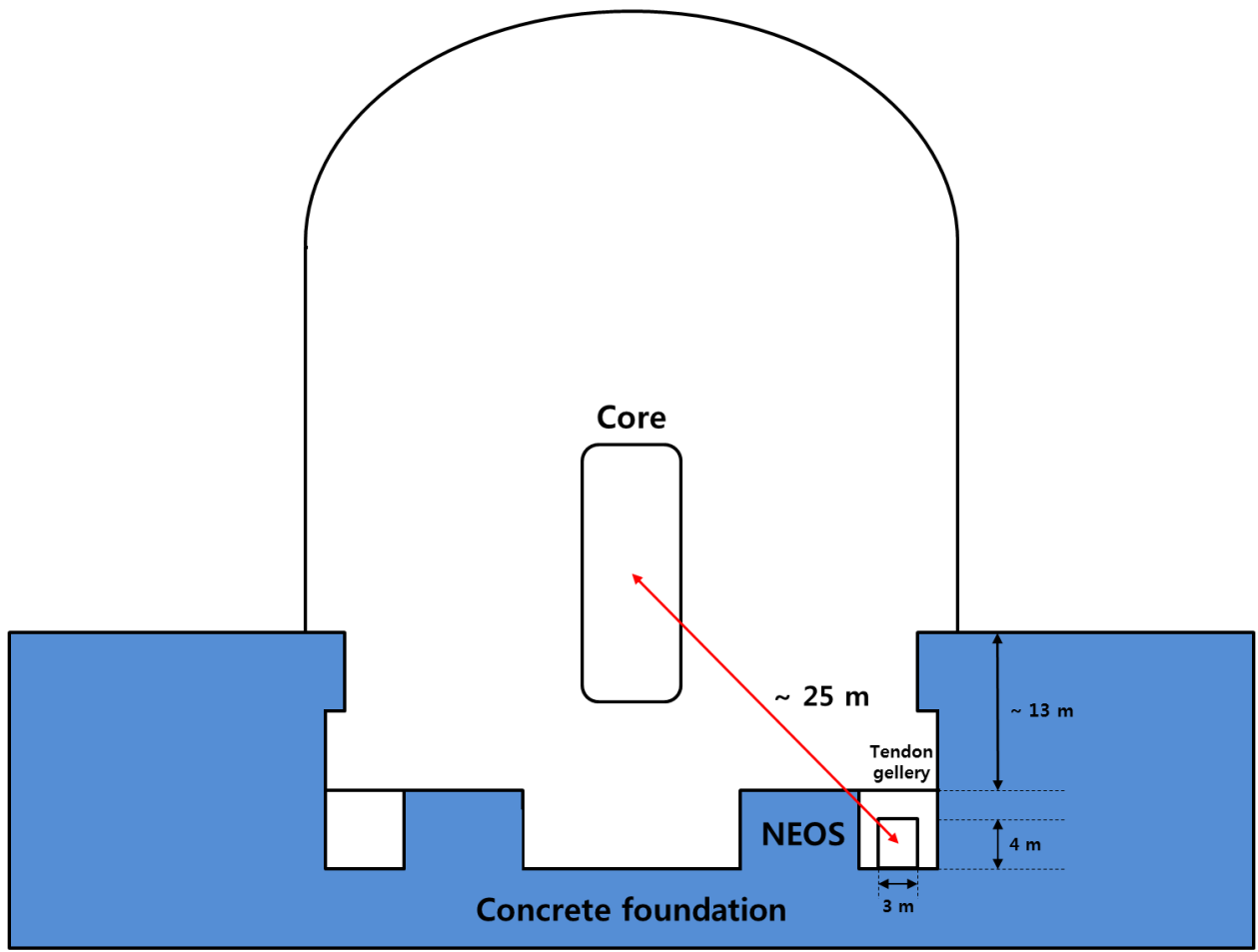

FIGURE 1: The NEOS experimental layout. The detector center located at the tendon gallery is approximately ~25 m away from the reactor core center. All dimensions not to scale.

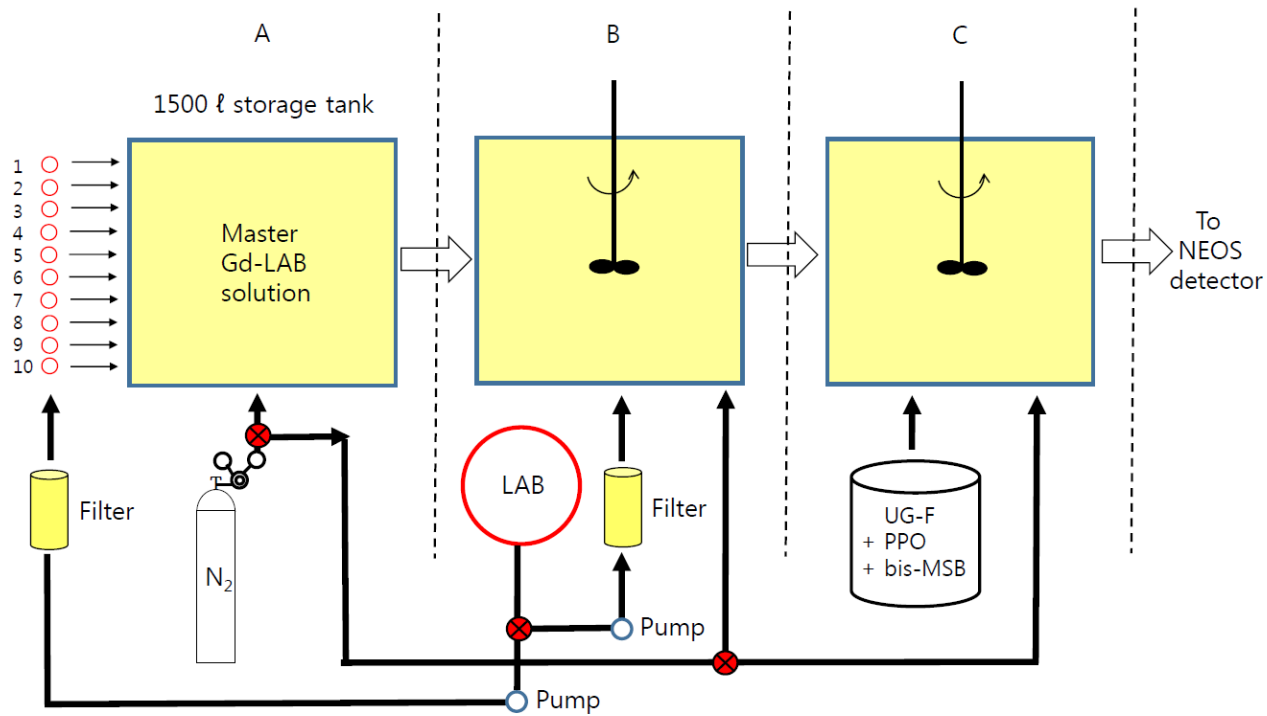

FIGURE 2: Three stages (A, B, and C) of the mixing scheme for the NEOS liquid scintillator. The size of mixing storage tank is 1500 ℓ. The $N_2$ gas and LAB flows are presented by solid lines and arrows.

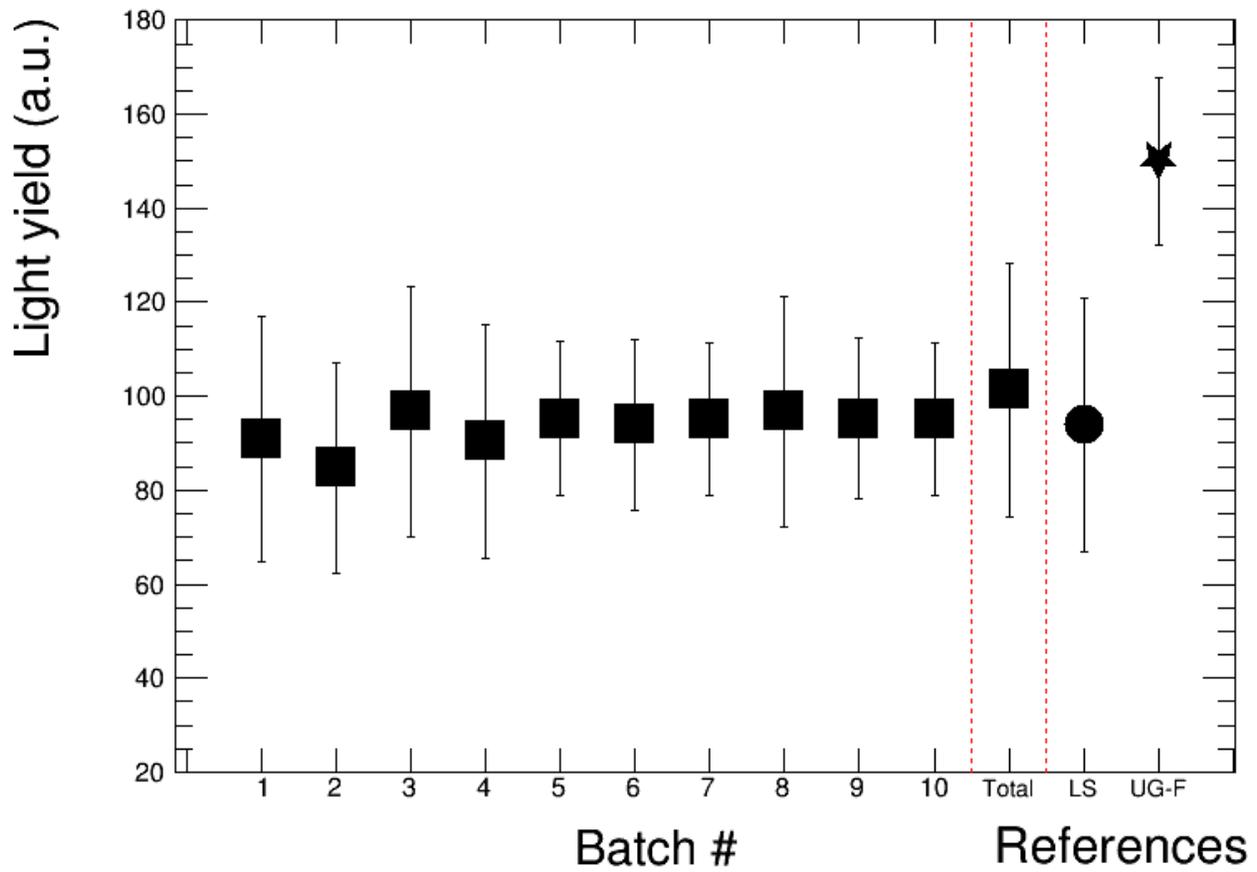

FIGURE 3: Light yield of ten batches in arbitrary units. UG-F (star mark) and LAB-based Gd-unloded LS (circle) are used as references for the comparison. The LY of ten GdLS samples is the same within the uncertainty. The LY of UG-F itself gives the highest value.

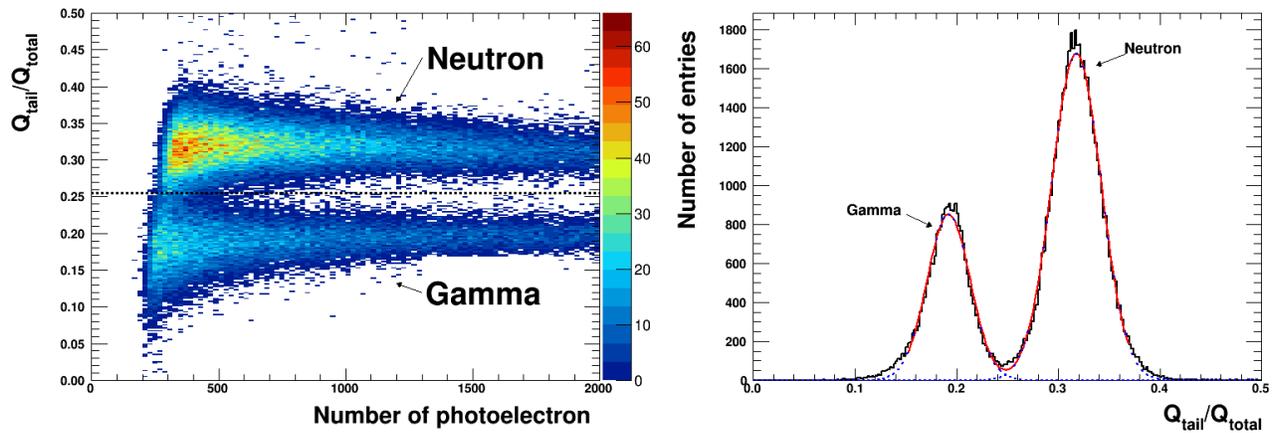

FIGURE 4: (Left) The scattered plot of the $\frac{Q_{\text{tail}}}{Q_{\text{total}}}$ as a function of the number of photoelectrons when the ratio of GdLS to UG-F is 90:10. Different responses to neutrons and gamma rays are shown. (Right) The number of entries as a function of $\frac{Q_{\text{tail}}}{Q_{\text{total}}}$ used to calculate the FoM. In the scatter plot, with a projection along the y-axis, it is possible to calculate the FoM value.

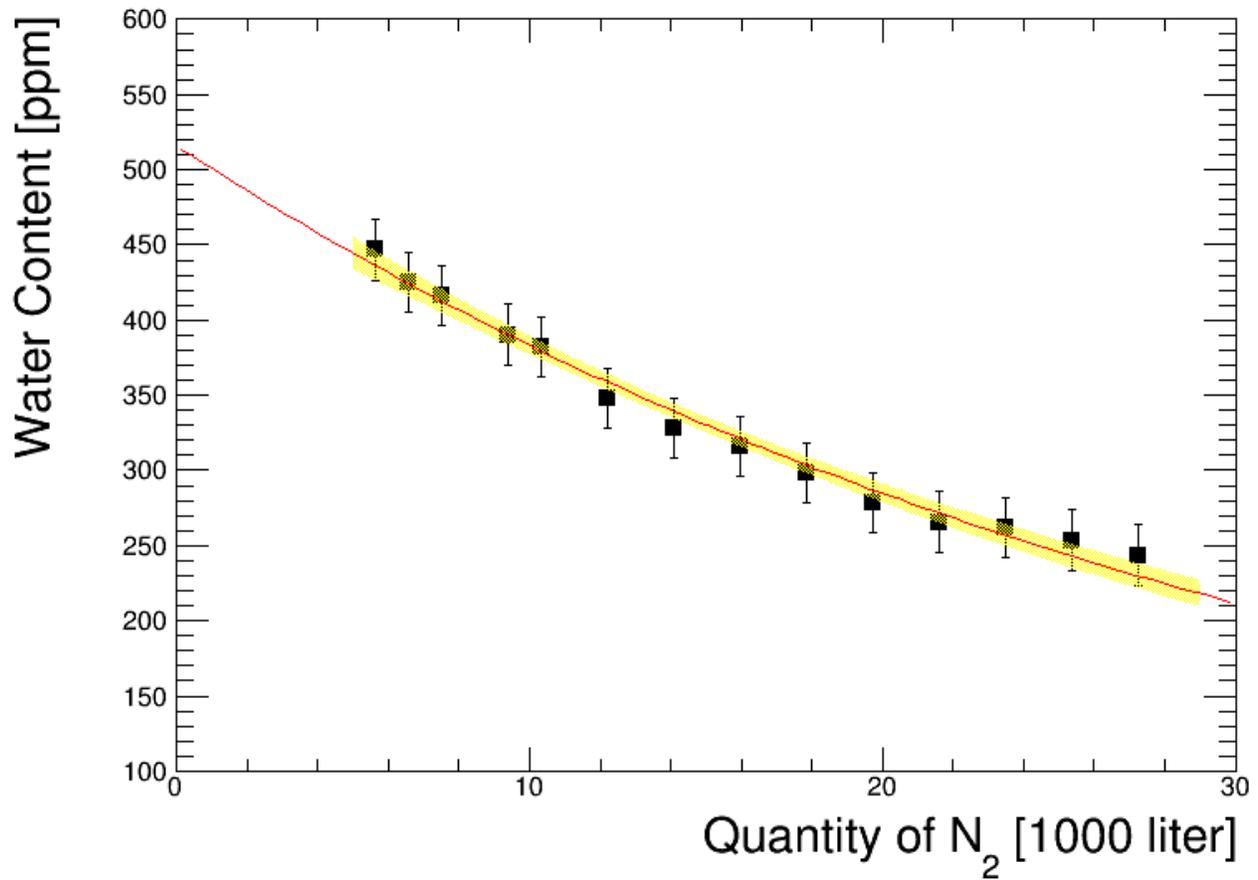

FIGURE 5: Water content as a function of the amount of $N_2$ gas injected into the GdLS. The water content decreases and then become stable at a value of ~200.0 ppm. All data points are fitted with a simple exponential function. The magnitude of the uncertainty is indicated by the band.

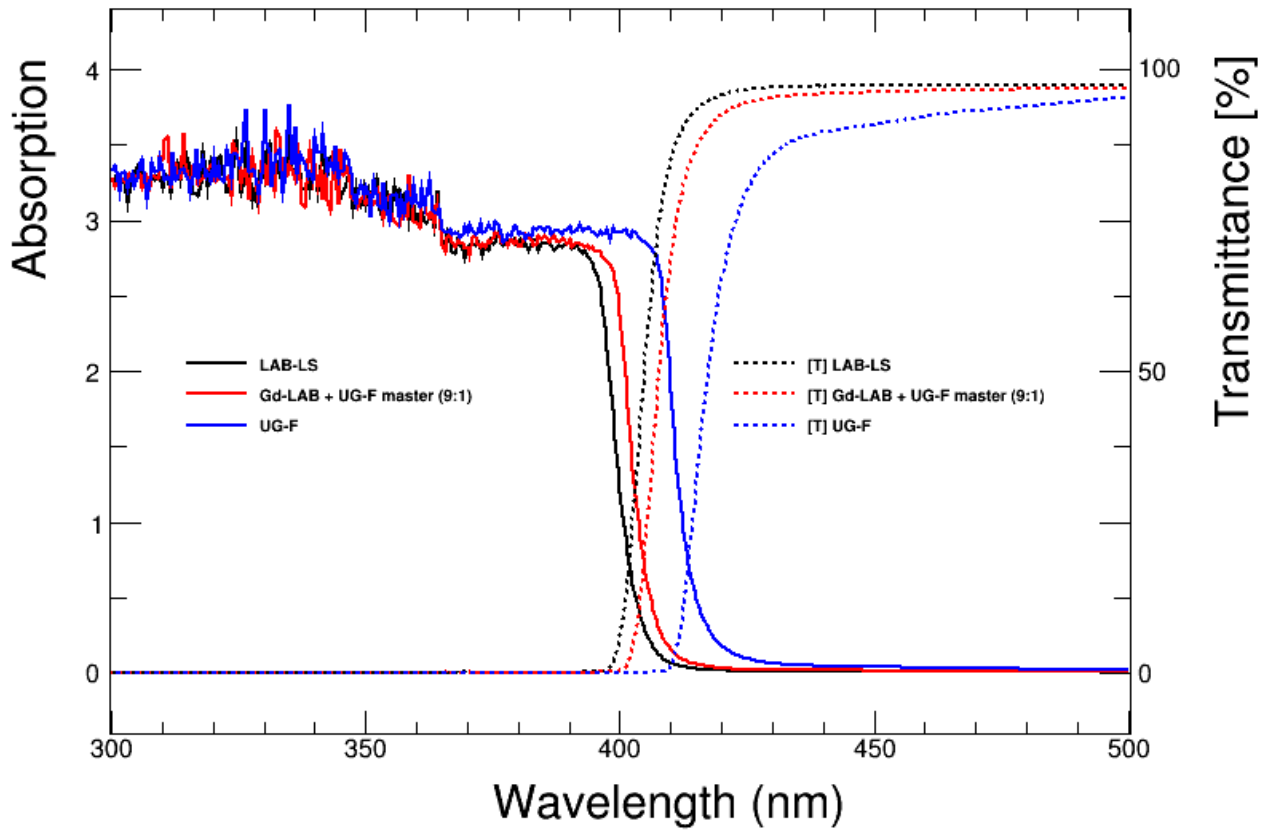

FIGURE 6: The transmittance (T) and absorption values of the NEOS liquid scintillator as a function of wavelength (300~500 nm) on one plot. The solid line represents the absorption, and the dashed line represents the transmittance. The transmittance value is given as a percentage. The averaged transmittance values of the LS and GdLS were over ~98% at the wavelength of 420 ~ 500 nm, where our PMTs are most sensitive and our wavelength region of interest lies. UG-F and LAB-based Gd-unloded LS are used for comparison.